# Broadband ferromagnetic resonance in Mn-doped Li ferrite nanoparticles

P. Hernández-Gómez[1*], J. M. Muñoz[1], M. A. Valente[2], M.P.F. Graça[2]

[1]Dpt. Electricidad y Electrónica, Univ. Valladolid, Paseo de Belén 7, 47011 Valladolid Spain

[2]Dpt. Fisica, Univ. Aveiro, Campus de Santiago, Aveiro, Portugal

*Corresponding author

e-mail: pabloher@ee.uva.es

Tel: +34983423895

**Abstract**

Lithium ferrites are well known materials due to their numerous technological applications especially in microwave devices. Mn-doped lithium ferrite nanoparticles were prepared by sol-gel technique by means of Pechini method, and then annealed at different temperatures in 250-1000º C range. XRD confirms spinel formation with particle size in the 15-200 nm range, with increased size with annealing temperature. Microwave magnetoabsorption data of annealed lithium ferrite nanoparticles, obtained with a broadband system based on a network analyzer operating up to 8.5 GHz are presented. At fields up to 200 mT we can observe a broad absorption peak that shifts to higher frequencies with magnetic field according to ferromagnetic resonance theory. The amplitude of absorption, up to 85 %, together with the frequency width of about 4.5 GHz makes this material suitable as wave absorber. Samples annealed at higher temperatures show a behaviour similar to polycrystalline samples, thus suggesting their multidomain character.

# 1. Introduction

Ferrimagnetic materials are widely used for electrotechnical equipment since their discovery in the forties. In particular, spinel ferrites are very good choices for magnetic recording, transformer cores and rod antennas due to their very high electrical resistivity, low eddy current losses, the possibility of tailoring their magnetic properties by changing composition, easy synthesis and cost production [1]. Lithium ferrite ($Li_{0.5}Fe_{2.5}O_4$) exhibits high Curie temperature, square hysteresis loop, moderate saturation magnetization and good thermal stability of their magnetic properties [2]. Due to the absence of divalent iron, it can also be employed in microwave devices, such as circulators, isolators, magnetostatic resonators, filters, switches, limiters and tunable electroptic modulators, replacing YIG for lower mass production costs [3]. Lithium ferrites are also good candidates in the field of rechargeable Li-ion batteries [4].

Lithium ferrite is a soft ferrimagnetic material with bcc cubic crystal lattice and magnetic ordering of inverse spinel structure, in which Li cations occupy octahedral B positions. In order to optimize their magnetic properties to the different microwave devices, appropriate chemical substitutions can be carried out, especially to reduce the losses. For this purpose it is known that small amounts of Mn added to Li ferrites decrease the dielectric loss, improve the remanence, and reduce the stress sensitivity of the remanence [5]. Moreover, Mn doping reduces porosity, grain size, magnetocrystalline anisotropy, magnetostriction, and also modifies the coercive field due to either displacement of Li cation to tetrahedral A sites or by reduction of some ferrous iron, this effect being especially remarkable in polycrystalline ferrites [6]. There are several studies that deal mainly with structural and dielectric properties of Mn substituted Li ferrites [7-10], employing different fabrication techniques different than conventional ceramic technique, because sintering at high temperatures promotes evaporation of lithium that made this material technologically difficult to prepare in this way. The sol-gel method provides an easy alternative for the preparation of nanosized lithium ferrites at lower annealing temperatures.

At present, due to exponential growth in microwave communication through mobile and satellite communications, there is strong interest in materials that absorb radio frequency energy. For wireless telecommunications, electronic measuring equipment, and to avoid interference noise in the frequency range up to 8 GHz, inexpensive, lightweight absorbers of electromagnetic radiation are needed. For broadband operation magnetic nanoparticles can be used [2]. Magnetic field induced microwave absorption in nanoscale ferrites is a recent and active area of research useful in this context [11]. In the present work, broadband microwave magnetoabsorption data of Mn substituted lithium ferrite nanoparticles are presented.

## 2. Experimental
### 2.1 Sample preparation

Mn-doped lithium ferrite nanoparticles were prepared by sol-gel technique by means of Pechini method. Starting materials were $LiNO_3$ (Fluka), $Fe(NO_3)_3 \cdot 9H_2O$ (Merck), $MnSO_4 \cdot H_2O$ ( Merck), citric acid ( Sigma Aldrich) and etilenglycol ( Fluka). Molar ratio among $LiNO_3$, $MnSO_4 \cdot H_2O$ and $Fe(NO_3)_3 \cdot 9H_2O$ was kept in 1:0.01:5 , so that spinel ferrite formation is optimized [6] with 1% mol Mn doping. Molar ratio for citric acid: metal was 3:1, and citric acid: etilenglycol was 1:2. In this way, nitrates and sulphates were solved in distilled water together with citric acid. Solutions were mixed with magnetic stirring for 30 min at room temperature. Then they were put together, mixed with etilenglycol and stirred for 1 h. to complete esterification reaction. Gel processing was achieved by drying at 90º C during 12h for water releasing, then heated at 150º C during 12h, and finally at 250º C for 1h. Gel volume grows indicating $NO_2$, $O_2$ and $CO_2$ releasing. Powders thus obtained were annealed at different temperatures in the 400-1000º C range. In all cases, the annealing procedure was carried out with 5 ºC/min heating rate and keeping the samples 4h at annealing temperature. This route of preparation has revealed to be one efficient and cheap technique to obtain high quality nanosized ferrite powder.

## 2.2 Measurement setup

X-ray diffractograms were obtained on a Siemens D5000 apparatus employing Cu-K$\alpha$ radiation ($\lambda$=1.54056 Å) at 40 kV and 30 mA with a curved graphite monochromator, an automatic divergence slit (irradiated length 20.00 mm), a progressive receiving slit (height 0.05 mm), and a flat plane sample holder in a Bragg-Brentano parafocusing optics configuration. Intensity data were collected by the step-counting method (step 0.02 º/s).

Magnetic field induced microwave absorption of nickel ferrite nanoparticles has been obtained with the help of an automatic measuring system based on a network analyzer Agilent model E5071C working from 0.1 MHz to 8.5 GHz. The sample holder is placed into the polar pieces of an electromagnet which produce magnetic fields up to 600 mT with a bipolar DC power supply Kepco BOP 50-8M. The magnetic field in the sample is measured with a gaussmeter FWBell 6010 with a calibrated perpendicular Hall probe. All the system is controlled with a PC with an appropriate Agilent VEE control program. Non-magnetic sample holder is placed at the end of a copper shorted semi-rigid coaxial line. The powdered sample is pressed into a toroidal shape that completely fills the space between the inner and outer conductors, which are short circuited at the end plane of the sample, ensuring that the sample is located in an area with minimum rf electric field and maximum rf magnetic field. Microwave absorption is obtained with the reflected rf signal by means of $S_{11}$ parameter, after translating the measurement plane to the sample position, and subtracting the signal obtained with the empty sample holder, so that we get only the absorption produced in the sample in the whole frequency range analyzed. This setup allow the broadband measurement of microwave absorption and hence the ferromagnetic resonance (FMR) with varying continuously both the operating frequency and DC magnetic field [12]. In addition, this measurement setup also allows the measurement of permittivity and permeability of small amounts of magnetic materials in the above mentioned frequency range with only the $S_{11}$ parameter [13].

## 3. Results and Discussion

In the figure 1 we can see that single phase spinel ferrite (JCPDS card 17-115) is obtained for all the annealing temperatures analyzed, except for the as-prepared powder. The average grain diameter has been obtained from them by using the Scherrer's formula for the [311] diffraction peak. Sample particle size increases with annealing temperature in good agreement with the increase in the sharpness of diffraction lines, related with the effect of annealing temperature on the higher crystallinity of the sample. Figures obtained with Scherrer's formula are the following: 15 nm for as prepared powders at 250º C, and 25 nm, 40 nm, 100 nm and 205 nm for samples annealed at 400º C, 600º C, 800º C and 1000º C resp. These values are similar to the findings by other authors [9, 14]. As expected, the sizes are smaller than Li ferrite samples with identical fabrication but without Mn doping (**due to the fact that similar sizes are expected for the as prepared samples after identical fabrication route, the size for the as prepared sample of Li ferrite is presented here as a reference, because XRD for the Mn doped sample cannot produce a representative value of particle size**) **[15].** With these average sizes superparamagnetic behaviour is not expected at room temperature. Higher annealing temperatures have not been analyzed, due to the formation of a secondary phase with lithium ferrate caused by the volatility of lithium [4].

In the figure 2, magnetic permeability data obtained with an LCR at 1 kHz are presented. We can see three different behaviours depending on the annealing temperature: samples annealed at 400º C- 800º C behave in a similar way, whereas powders obtained at 250º C without annealing are non-magnetic due to incomplete formation of ferrimagnetic spinel ferrite. Finally, the sample at 1000º C has a higher relative magnetic permeability regarding the rest of samples. As we will discuss below, it is probably related to the existence of magnetic domains in this sample. The absence of local maxima in these curves suggests that the ferrimagnetic character of the samples remains in the temperature range 80 K - 420 K.

As a reference of the broadband nature of experimental results, we show in the figure 3a a 3D plot of the microwave absorption. Similar information can be displayed in colour 2D plots of absorption vs magnetic field or absorption vs frequency. In the figure 3b we can see the linear FMR behaviour of the sample annealed at 400º C. Additionally the usual FMR curves of derivative of absorption vs magnetic field at a fixed frequency could also be prepared.

In the figure 4 we present the results of microwave magnetoabsorption curves of the different thermally annealed Mn doped lithium ferrites (the as-prepared sample at 250º C is not shown as it doesn't exhibit microwave absorption). In each graph the frequency behaviour of the measured reflected rf signal for some selected values of the applied DC magnetic field is shown. In all the curves we observe a single peak of maximum absorption that shifts to higher frequencies with increasing the applied magnetic field. In the frequency range available by our measurement setup, we can observe the absorption peaks at applied fields up to 200 mT. Higher fields shift the peak to frequencies beyond the capability of our network analyzer. These peaks are very broad in frequency (up to 4 GHz half width in sample sintered at 800º C), so that they could be good candidates to microwave absorbers. In addition, the sample annealed at 1000º C also exhibits a secondary process near the higher frequency range available by our system. The behaviour of this sample is rather similar to the results obtained in a sample of polycrystalline Li ferrite prepared in similar conditions, which is also presented for comparison. It is also noteworthy that the amount of sample needed to fill the sample holder is similar in these last two samples, about 100 mg, and four times higher than the rest, pointing to a higher densification of the sample annealed at 1000º C regarding the other samples annealed at lower temperatures.

Concerning the strength of microwave absorption, we can establish two different behaviours: samples sintered up to 800º C have similar figures of about 40 % absorption (about 2% per gram). On the other hand, in the sample annealed at 1000º C the absorption increases up to 85% (i.e. 0.85% per gram), with a qualitatively different behaviour, because the maximum absorption is obtained without applied field, whereas in the rest of the samples the absorption increases with the applied field. The figure of maximum absorption is also closely related to the result obtained with bulk Li ferrite. The fact that the results of magnetoabsorption qualitatively and quantitavely match, point to the possibility that sample annealed at 1000º C contains multiple magnetic domains. This fact is supported by the average crystallite size (205 nm), as observed in other magnetic systems [12], the higher density of the obtained compound, comparable to polycrystalline ferrite, and the higher magnetic permeability values as shown in Figure 1.

These results resemble to previous ones obtained by us and other authors [12, 16], and can be ascribed to ferromagnetic resonance of the spinel ferrite. In order to obtain the characteristic parameters of FMR, the resonance frequency as a function of the magnetic applied field is represented in the figure 5. Data for samples annealed at 1000º C and bulk ferrite are difficult to extract due to overlapping of two phenomena: thermal annealing enhance crystallite growth and promotes multidomain structure which allow spin as well as domain wall resonances, that broaden and distort the resonance line at lower frequencies (this effect can also be present to some extent in the sample annealed at 800º C). In addition, the secondary process near our top frequency range also affects the resonance data. With this in mind, we have tried to fit the data to the Kittel expression for FMR in spheres:

$$f_r = \frac{g \cdot \gamma}{2\pi} (B_r + B_A) \tag{1}$$

with $f_r$ the resonance frequency, $B_r$ the applied magnetic field at resonance, g is the spectroscopic splitting factor, $\gamma$ is the gyromagnetic ratio, and $B_A$ the effective anisotropy field. It is noteworthy that in this frequency range, the anisotropy field cannot be ignored, as it is similar, or even higher than the applied magnetic field. The results are not conclusive, but we can extract g-values around 2.18, similar to the 2.139-2.158 range found in literature data for bulk ferrites [17]. Anisotropy fields lie in the range from 0.9 kOe for sample annealed at 400º C to 3.30-3.70 kOe for the rest of the samples (values have been converted to H in cgs units to allow comparison with literature data), similar to the figures obtained previously for other nanoparticulate spinel ferrite systems [11, 12], but one order of magnitude higher than bulk Li ferrites [18].

In order to obtain additional parameters to ascertain the validity of our results, we have measured the magnetic parameters of the sample annealed at 1000º C, obtaining a saturation magnetization of 57 emu/g and a coercive field of 250 Oe. We observe that the coercive field is fairly lower than the calculated anisotropy field, as it happens in multidomain compounds. $M_S$ is lower than bulk magnetization, thus suggesting spin disordering in the surface. We can make a rough estimation of the shell thickness with the expression

$$M_S(D) = M_{SBulk}\left(1 - 6\frac{t}{D}\right) \tag{2}$$

obtaining a thickness around 5 nm, i.e. 6 lattice steps of surface layer, so that we can model our samples as a core-shell assembly of interacting particles. The fact that $M_S$ is 10 % higher than the non-doped Li ferrite is not surprising, because the Mn doping in Li ferrite polycrystals promote a diminution of ferrous iron content that appears in the annealing process [6]. In this case we can also consider that Mn cations enter into the octahedral sites in the spinel structure [8, 19], displacing Li atoms to tetrahedral sites as well as to the chemically disordered surface layer, thus promoting the observed magnetization increase regarding undoped Li ferrites.

With the measured $M_S$ and the obtained anisotropy field after FMR fitting, we can calculate the value of the effective anisotropy constant, to get $3.75 \cdot 10^4 \pm 0.55 \cdot 10^4$ J/m$^3$ for the sample annealed at 1000º C. This value is four times higher than reported for bulk lithium ferrites. It is noteworthy that for magnetic nanoparticles the value of the effective anisotropy constant is determined not only by the contribution of the bulk magnetocrystalline anisotropy, but also by the surface, strain and shape anisotropy, as well as the anisotropy arising from interparticle interactions, and in our case the value obtained is very close to the value of $4 \cdot 10^4$ J/m$^3$ obtained with a different measurement approach by Wang in a very similar ferrite system [14] assuming a core-shell configuration of the nanoparticles annealed at lower temperatures, and multidomain behaviour in the higher size particles, in a similar way as our findings.

With this in mind, we can consider the effect of Mn on the microwave absorption of nanoparticle lithium ferrites. It is well known that in polycrystals, Mn substitution prevent the grain growth and reduce porosity. Our results confirm that average nanoparticle size is lower than undoped samples annealed at the same temperature [15].

In addition, Mn cations can impede the existence of ferrous cations by appropriate valence change, and then $Mn^{2+}$ and some amount of $Mn^{3+}$ are present. $Mn^{2+}$ has low influence in anisotropy, but $Mn^{3+}$ is a Jahn Teller cation that can alter the local crystal fields, and hence the anisotropy constant [6]. It is also known that magnetostriction constants are strongly lowered by Mn addition in Li ferrites [5]. The increase in surface to volume ratio produces a non negligible shell in which the broken bonds produce a different chemical and exchange effect than in the core. In this ferrite system some Li displacement takes place from the core to the shell [19]. The overall effect of the above mentioned factors is an increase in surface anisotropy than can exceed the bulk anisotropy up to two orders of magnitude. The increased surface anisotropy and the inhomogeneities in the shell broaden the resonance [6]. Preliminary measurements show that the resonance linewidth is 10% higher in Mn doped Li ferrite nanoparticles than the non doped samples prepared with the same annealing temperature [15]. Finally, the observed variation in the microwave absorption results of the analyzed samples annealed at different temperatures arise mainly from the different particle sizes, because the Mn content do not change, together with the previously mentioned multidomain formation at higher annealing temperatures. For lower annealing temperatures, the lower particle size promotes a higher anisotropy and surface to volume ratio that causes a broader and higher absorption.

## 4. Conclusions

Magnetic field induced microwave absorption at frequencies up to 8.5 GHz have been measured in different thermally annealed Li ferrite nanoparticles obtained with sol-gel Pechini method. The results agree well with the theory of ferromagnetic resonance. The wide microwave absorption observed reveal that this material can be used as absorber in this frequency band. Nanoparticle Li ferrites have higher effective anisotropy fields than bulk ferrites, caused by the increased contribution to the effective anisotropy due to the surface inhomogeneity.

**Acknowledgements**

Funding: This work was supported by the Spanish Ministerio de Economía, Industria y Competitividad, Agencia Estatal de Investigación with FEDER, project id. MAT2016-80784-P

**Figure Captions**

Figure 1 X ray diffractograms of the as prepared and annealed Mn doped Li ferrite samples.

Figure 2. Magnetic permeability of the annealed Mn doped Li ferrite nanoparticles.

Figure 3 a) 3D plot of magnetoabsorption vs applied magnetic field and frequency corresponding to sample annealed at 400º C. b) 2D colour map of absorption at different applied magnetic field vs frequency.

Figure 4. Microwave absorption as a function of frequency, at different applied magnetic fields (curves shift with increasing applied field from left to right in each graph), corresponding to Mn doped $LiFe_{2.5}O_4$ nanoparticle samples annealed at 400º C, 800º C, and 1000º C, as well as bulk ferrite sample.

Figure 5. Resonance frequency as a function of applied magnetic field, corresponding to Mn doped Li ferrite nanoparticles annealed at 400-1000º C, together with bulk ferrite.

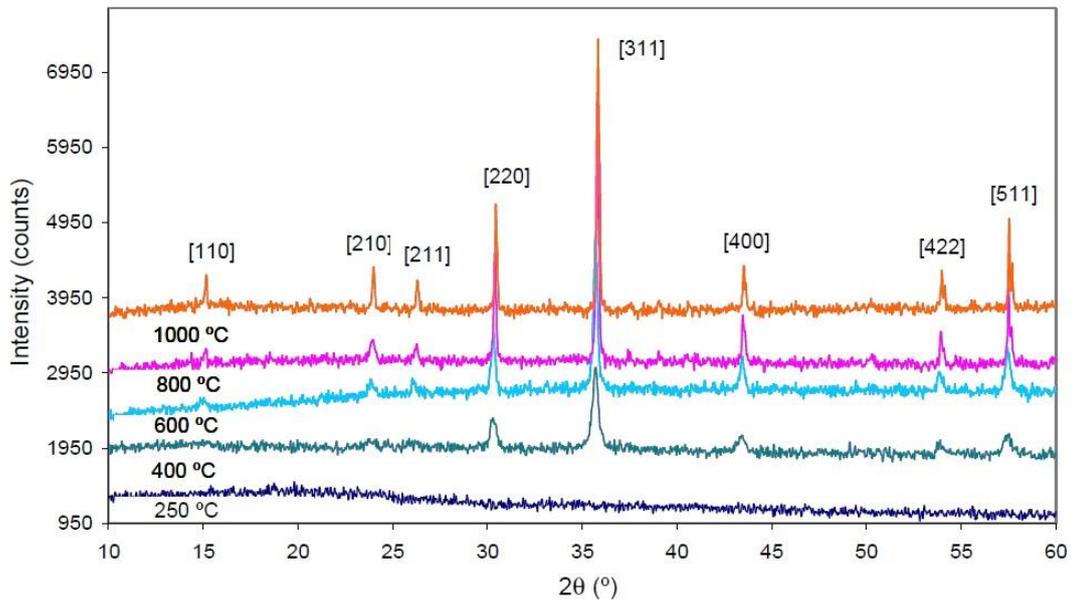

Figure 1 X ray diffractograms of the as prepared and annealed Mn doped Li ferrite samples.

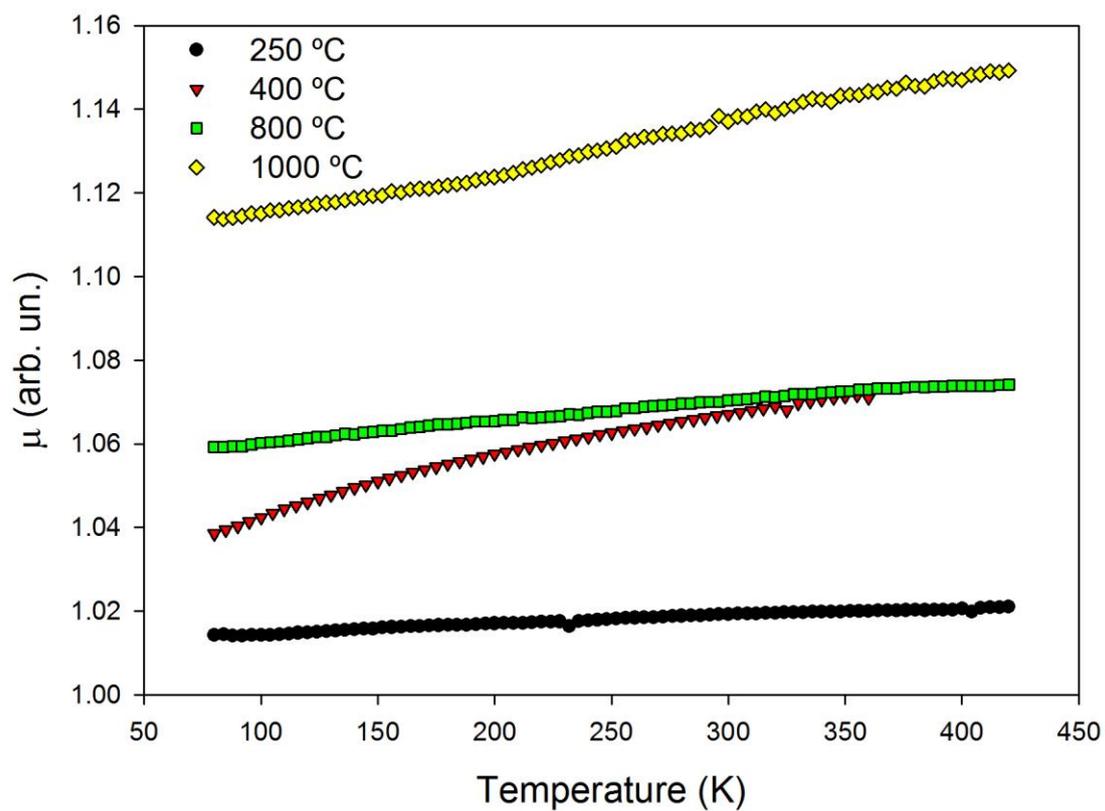

Figure 2. Magnetic permeability of the annealed Mn doped Li ferrite nanoparticles.

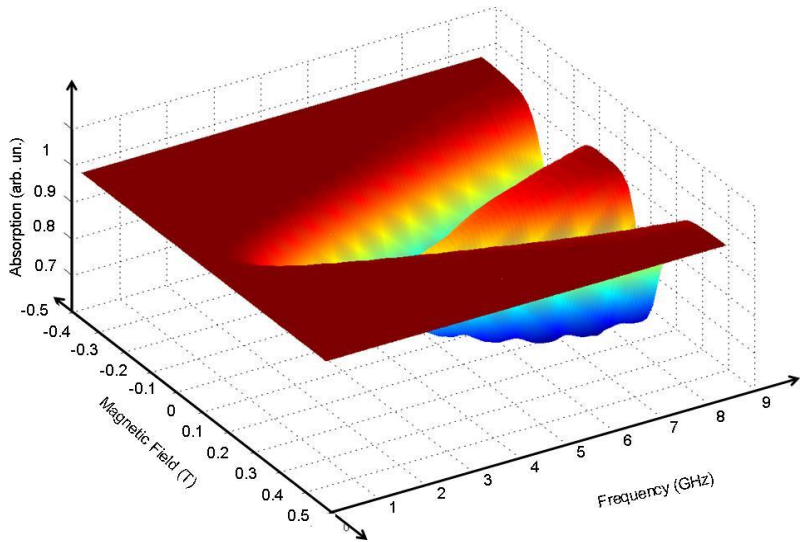

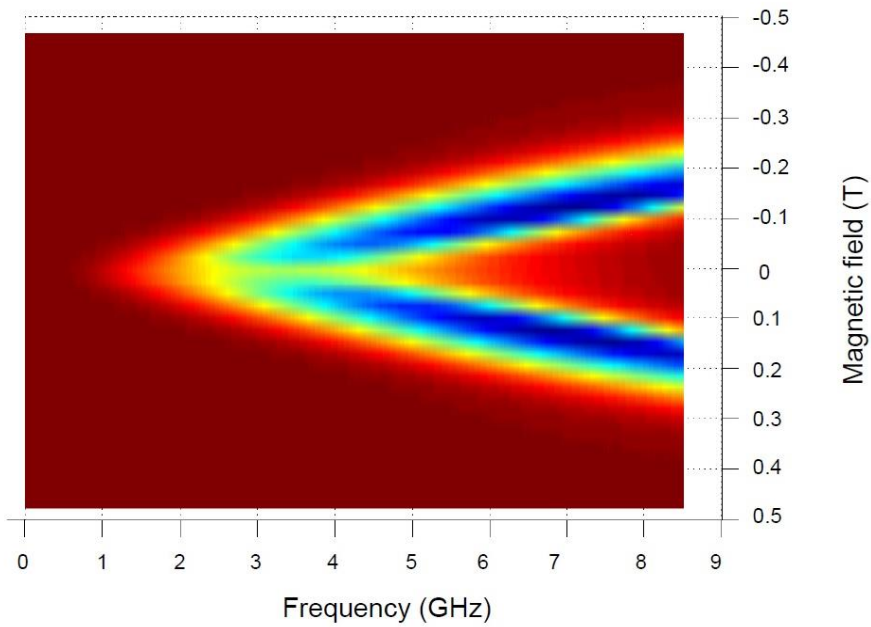

Figure 3 a) 3D plot of magnetoabsorption vs applied magnetic field and frequency corresponding to sample annealed at 400º C. b) 2D colour map of absorption at different applied magnetic field vs frequency.

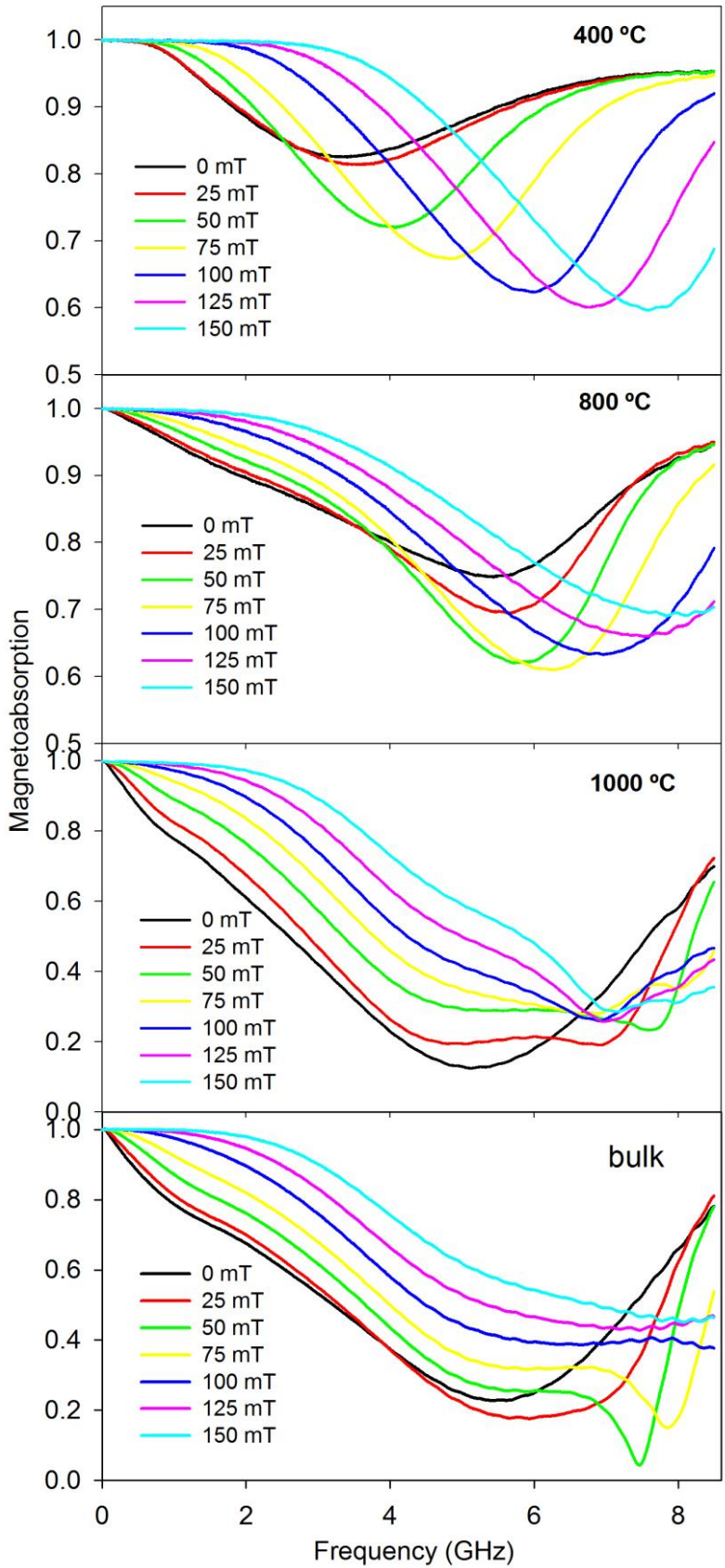

Figure 4. Microwave absorption as a function of frequency, at different applied magnetic fields (curves shift with increasing applied field from left to right in each graph), corresponding to Mn doped LiFe$_{2.5}$O$_4$ nanoparticle samples annealed at 400º C, 800º C, and 1000º C, as well as bulk ferrite sample.

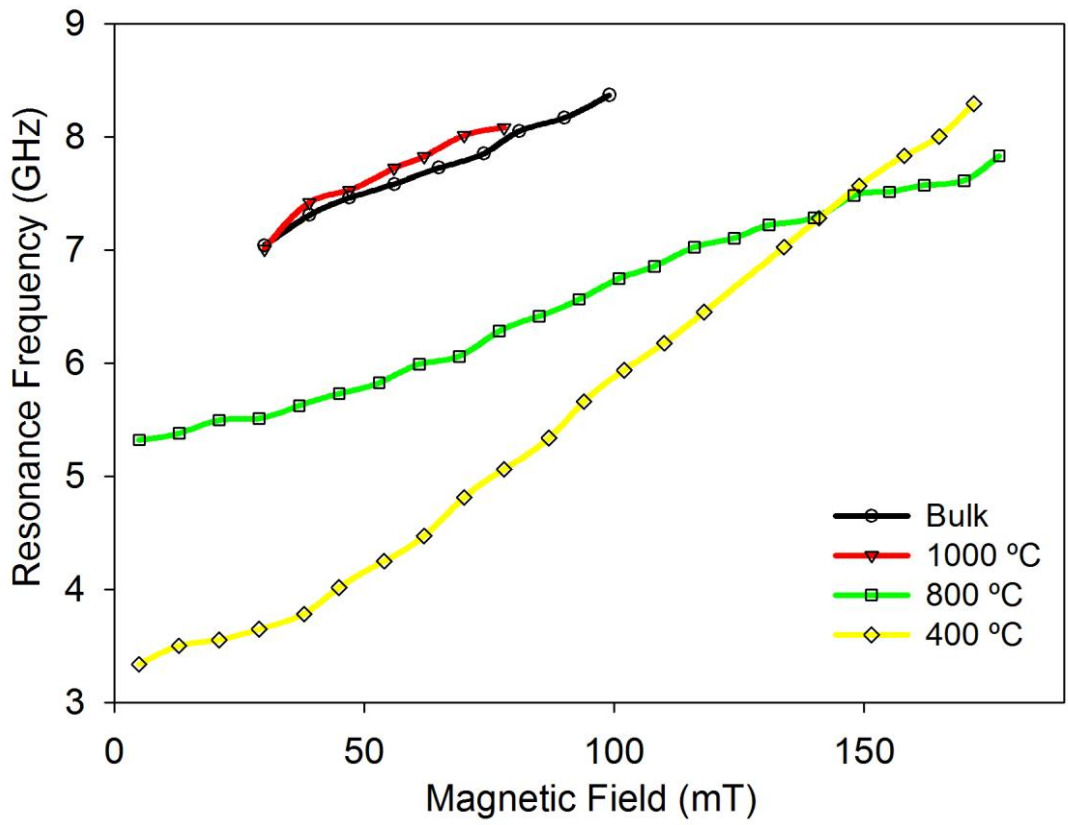

Figure 5. Resonance frequency as a function of applied magnetic field, corresponding to Mn doped Li ferrite nanoparticles annealed at 400-1000º C, together with bulk ferrite.